\documentclass[prl,aps,twocolumn,showpacs,superscriptaddress]{revtex4}
\usepackage{bm}
\usepackage{amsmath,amssymb}
\usepackage{graphicx}
\usepackage{dcolumn}

\begin{document}

\title{Amine-Gold Linked Single-Molecule Junctions: Experiment and Theory}
\author{Su Ying Quek}
\affiliation{Molecular Foundry, Lawrence Berkeley National Laboratory, Berkeley, CA}
\author{Latha Venkataraman}
\affiliation{Department of Physics, Columbia University, New York, NY}
\affiliation{Center for Electron Transport in Nanostructures, Columbia University, New York, NY}
\author{Hyoung Joon Choi}
\affiliation{Department of Physics and IPAP, Yonsei University, Seoul, Korea}
\author{Steven G. Louie}
\affiliation{Molecular Foundry, Lawrence Berkeley National Laboratory, Berkeley, CA}
\affiliation{Department of Physics, University of California, Berkeley, CA }
\author{Mark S. Hybertsen}
\affiliation{Center for Functional Nanomaterials, Brookhaven National Laboratory, Upton, NY}
\affiliation{Center for Electron Transport in Nanostructures, Columbia University, New York, NY}
\author{J.B. Neaton}
\email{JBNeaton@lbl.gov}
\affiliation{Molecular Foundry, Lawrence Berkeley National Laboratory, Berkeley, CA}

\begin{abstract}
The measured conductance distribution for single molecule benzenediamine-gold junctions, based on 59,000 individual conductance traces recorded while breaking a gold point contact in solution, has a clear peak at 0.0064 G$_{0}$ with a width of $\pm$ 40\%.  Conductance calculations based on density functional theory (DFT) for 15 distinct junction geometries show a similar spread.  Differences in local structure have a limited influence on conductance because the amine-Au bonding motif is well-defined and flexible. The average calculated conductance (0.046 G$_{0}$) is seven times larger than experiment, suggesting the importance of many-electron corrections beyond DFT.
\end{abstract}

\date{June 12, 2007}
\maketitle

Discovering the anatomy of single-molecule electronic circuits, in order to exploit their transport behavior, poses fundamental challenges to nanoscience \cite{Nitzan03}. Single-molecule manipulation requires probes at the limits of experimental resolution \cite{Reed97,Reichert02,Liang02,Xu03,Venk06,Moresco01}, as control at the single-bond scale is necessary for reliable transport properties. The success of amine-Au links for the realization of single-molecule junctions has allowed systematic measurements of the relationship between molecular-scale structure and conductance for several families of molecules \cite{Venk06,Venk06nano,Venk07}. Experimental trends were explained based on the hypothesis of selective bonding between the amine lone pair and an undercoordinated gold atom \cite{Venk06nano}. Understanding the essential junction features that lead to reproducible electrical properties would enable generalization to other metal-molecule link chemistries, while also furthering the development of a predictive theoretical approach to nanoscale transport. 

Most recent theoretical investigations of nanoscale transport rely on a Landauer approach, simplified to treat electronic interactions at a mean-field level within density functional theory (DFT). While this framework has proven relatively accurate for metallic point contacts \cite{Nielsen02}, the computed conductance often substantially exceeds measured values for molecular junctions \cite{Reed97,Basch05,Stokbro03,Tomfohr04,Choi07}, raising questions about fundamental theoretical approximations. However, since the predicted conductance can be very sensitive to contact geometry (changing by orders of magnitude for thiol-Au junctions, for instance \cite{Stokbro03,Tomfohr04,Muller06}), uncertainty over junction structure has obscured this issue. A clear benchmark is required. 

In this Letter, we use the reproducible conductance of amine-gold linked single-molecule circuits \cite{Venk06,Venk06nano,Venk07} to develop a detailed, atomic-scale understanding of a prototypical molecular junction. A large experimental data set for benzenediamine (BDA) is analyzed in depth, and directly compared with DFT-based transport calculations.  We show that amine-gold bonding is highly selective, but flexible, resulting in a conductance insensitive to junction structure. The computed conductance has a narrow spread consistent with experiment, although the average calculated conductance is $\sim$7$\times$ larger. Our benchmark comparison demonstrates the importance of many-electron corrections beyond DFT.

\begin{figure}[top]
  \includegraphics[width=6.8cm]{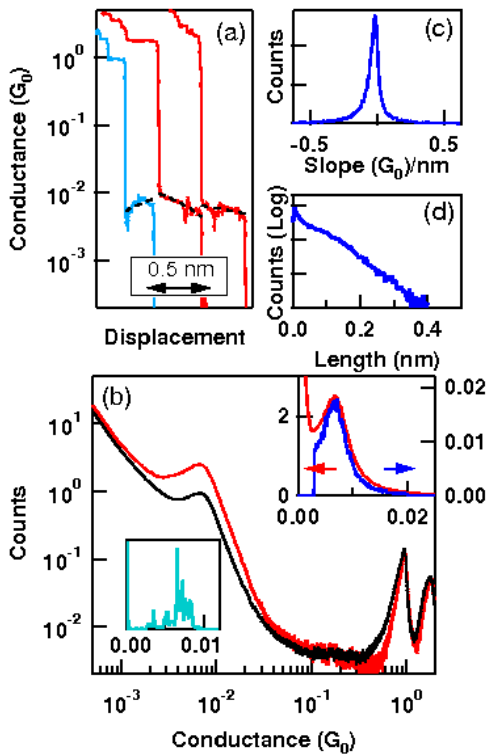}
\caption{
\label{fig1}(color online) (a) Sample conductance traces. Dashed lines are linear fits to the molecular step determined by an automated algorithm (applied bias: 25 mV). (b) Conductance histograms of all 59000 measured traces (black) and of 21312 selected traces with steps (red), shown on a log-log scale.  Inset (top-right): Conductance histogram of selected traces (red) and histogram of average step conductance (blue). Inset (center): Histogram of the blue conductance trace in (a) computed by binning the conductance data (acquired at 40 kHz). All histograms in (b) are obtained using a 10$^{-4}$ G$_{0}$ bin size and are normalized to the number of traces. (c-d) Normalized histograms of (c) step lengths (bin size: 0.004 G$_{0}$/nm) and (d) step slopes (bin size: 0.0025 G$_{0}$/nm).}
\end{figure}

To obtain a detailed picture of junction structure that can be connected to its transport properties, we analyze the distribution of low-bias conductance of Au-BDA-Au junctions measured by repeatedly breaking Au point contacts in a molecular solution \cite{Venk06}. Many conductance traces reveal a step below 1 G$_{0}$ ($2e^{2}/h$)  (Fig. 1(a)), corresponding to a single BDA molecule between the two atomic point-contacts \cite{Venk06,Venk06nano}. Fig. 1(b) shows the histogram (black) computed from all 59,000 measured traces, from 4 different tip/sample pairs, without any data selection. The most probable junction conductance obtained from Lorentzian \cite{Venk06} fits to the histogram peak is 0.0064 $\pm$ 0.0005 G$_{0}$, with half width 47 $\pm$ 8 \%. 

Not all measured conductance traces have a step at a value near the molecular peak in the full histogram (black, Fig. 1(b)). Applying an automated sorting algorithm \cite{algo} to the full data set indicates that a molecule is trapped between the breaking Au contacts in about 36\% of the traces. Our algorithm also determines the length and slopes of the molecular steps, whose distributions are shown in Fig. 1(c-d). We find that the majority of steps ($\sim$75\%) are $<$ 1 $\text{\AA}$ long, with a power law distribution extending to about 4 $\text{\AA}$. Further, $\sim$75\% of the steps are flat, with a slope ranging from -0.15 to -0.05 G$_{0}$/nm.

Further insight is gained from comparing the distribution of step-averaged conductances in traces with a molecular step (blue histogram in Fig. 1(b) inset) to that obtained by binning all data points in the traces (red histogram in Fig. 1(b) inset). The conductance peak position and half width are 0.0065 (0.0065) G$_{0}$ and $\pm$ 40 (45) \% for the blue (red) histograms, in excellent agreement with those for the full histogram (black). The slightly narrower peak width in the step-average histogram indicates the peak width (red) originates primarily from variations in conductance among different junctions: Junction-elongation in individual traces evidently does not significantly add to the spread in conductance.  

The repeated plastic deformation of the contact region from measurement to measurement induces variations in junction structure. Understanding the impact of junction structure on conductance is critical. We approach this problem in two steps.  Using first-principles DFT, possible bonding motifs in the junction are analyzed to probe the selective bonding hypothesis \cite{Venk06nano}. Then, guided by allowed bonding motifs, 15 distinct junction geometries are examined to relate the conductance to junction structure. 

Our DFT calculations are performed within the generalized gradient approximation (GGA) \cite{Perdew96}, as implemented in the SIESTA code \cite{Soler02}. An optimized single-$\zeta$ basis set is used for the Au $d$ shell; all other orbitals are described by double-$\zeta$ polarization basis sets \cite{Au}. Junctions are first constructed with 6-layer Au(111) slabs on either side of the BDA molecules. A supercell geometry with 16 Au atoms per layer is used. All atoms in the junction, except for those in the bottom 3 Au layers of each slab, are relaxed until forces are $<$ 0.05 eV/$\text{\AA}$. The conductance is obtained using a coherent scattering-state approach based on DFT \cite{Choi07}. The BDA junction is divided into three regions: left bulk, center resistive region (including 4 Au layers on either side of molecule), and right bulk. The Hartree potential in the center region smoothly matches that of the bulk regions, which are infinitely repeated away from the junction to simulate open boundary conditions. The two-dimensional Brillouin Zone is sampled by an 8 $\times$ 8 Monkhorst-Pack $k_{\parallel}$-mesh in most junctions; in the (III$^{\prime}$, III$^{\prime}$) junctions, a 16 $\times$ 16 mesh is found to be necessary for converged conductance. Energy- and $k_{\parallel}$-dependent scattering states are constructed with incoming and outgoing states determined from the bulk Au complex band structure. The zero-bias conductance is computed from the Landauer formula G = G$_{0}$ $\Sigma_{k_{\parallel}}$ Tr($t^{\dagger}t$), where the transmission matrix $t$ is evaluated at the Fermi energy, $E_{F}$. 

Unlike thiol-Au linkages \cite{Tomfohr04}, we find that the local amine-Au bonding motif is remarkably well-defined, with the amine group only binding to undercoordinated Au atop sites. If the amine group is initially positioned at the bridge or three-fold site between Au atoms, the molecule is found to relax to the nearest atop site, or drift away from the Au contact. Furthermore, typical BDA binding energies on undercoordinated Au sites (-0.4 to -0.6 eV) are significantly larger than that on the atop site of atomically flat Au(111) (-0.1 eV), in accord with previous DFT calculations on alkanediamines \cite{Venk06nano}. This provides detailed support for the initial hypothesis \cite{Venk06nano}: the strongly basic amine group readily donates its lone pair to form Au-N bonds with an undercoordinated Au atom where the 6$s$ orbital is more available, in contrast to the relatively unreactive sites on flat Au(111) \cite{Venk06nano}.

\begin{figure}[top]
  \includegraphics[width=6.6cm]{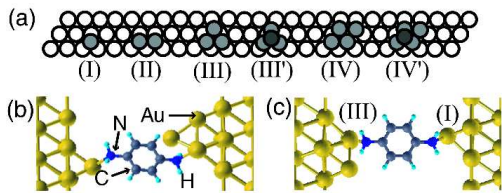}
\caption{
\label{fig2} (a) Au contact motifs, (b) 'side' junction, (c) (III, I) junction. In (b), one amine group is bonded to an adatom (I), and the other to an atom at the base of a pyramid (III$^{\prime}$).}
\end{figure}

We next develop model contact geometries with variations in junction structure, considering a total of 15 fully-relaxed molecular junctions.  These geometries differ in the molecular tilt angle (0$^\circ$, 10$^\circ$, 31$^\circ$, 56$^\circ$ with Au surface normal), bonding configuration (\textit{cis}/\textit{trans}), binding sites, and Au contact structures.  Both symmetric and asymmetric junctions are examined. Six distinct Au contact structures with varying degrees of coordination are considered (Fig. 2(a)). Except for the 'side' junction in Fig. 2(b), all other junctions are labeled according to its two Au contact structures (e.g., (III, I) in Fig. 2(c)). The calculated binding energy for the BDA varies modestly among the 15 junction structures (0.8-1.3 eV). 

Fig. 3 shows a typical calculated transmission curve, $T(E)$, at zero bias, obtained for the (III, I) junction. The peaks centered at -1.2, 2.7 and 3.2 eV derive respectively from the BDA highest occupied molecular orbital (HOMO), lowest unoccupied molecular orbital (LUMO) and LUMO + 1 states. The transmission for all 15 junctions is plotted on a log-scale in Fig. 4(a), and illustrates the remarkably narrow spread in $T(E)$  near  $E_{F}$. Moreover, for all junctions, the transmission from $\sim$ -1.5 to 0.5 eV is found to be associated with a single eigenchannel \cite{Brandbyge97} arising from the BDA HOMO, consistent with experiment \cite{Venk07}.  For the 15 junctions considered here, the calculated linear response conductance G$_{0} T(E_{F})$ has a mean value of 0.046 G$_{0}$, and ranges from 0.032 (-31.1 \%) to 0.062 (+35.4 \%) G$_{0}$.  Notably, the spread is similar to the half-width observed in the histogram of measured step-average conductances (Inset, Fig. 1(b)); the calculated conductance magnitude, however, exceeds the measured value by about a factor of seven.

\begin{figure}[top]
  \includegraphics[width=5.5cm]{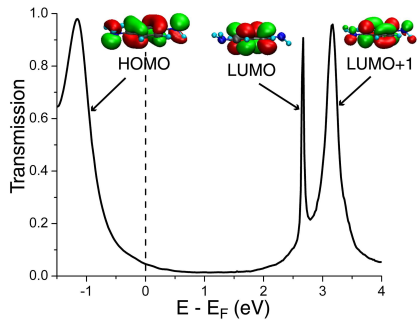}
\caption{
\label{fig3} Energy-dependent transmission for the (III, I) junction. Isosurface plots illustrate the molecular orbitals responsible for the transmission peaks.}
\end{figure}

\begin{figure}[top]
  \includegraphics[width=5.6cm]{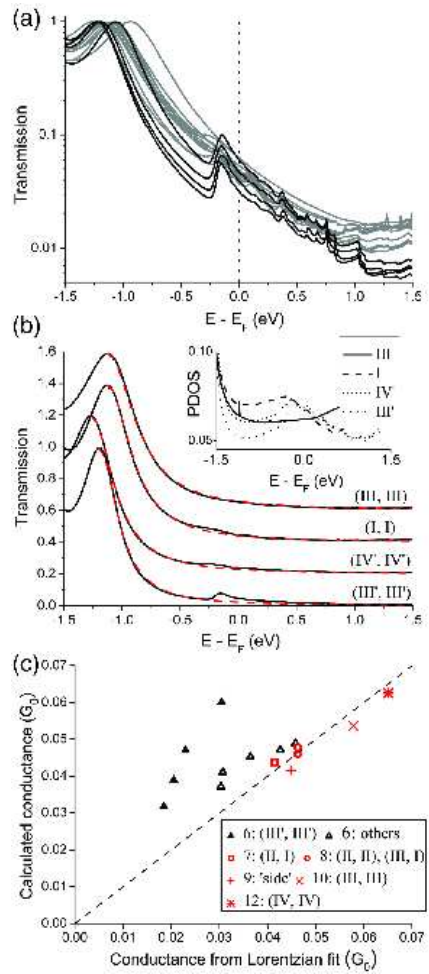}
\caption{
\label{fig4} Calculated transmission on (a) a log scale for all 15 junctions (black: (III$^{\prime}$, III$^{\prime}$) junctions), and (b) a linear scale for 4 representative geometries, vertically offset for clarity. Inset: Scattering state PDOS on Au binding sites, in the absence of BDA. (c) Calculated conductance versus conductance predicted from Lorentzian fits (red dashed lines in (b)). The resonances centered at $\sim$ -1 eV have a maximum transmission of 0.97-0.99 G$_{0}$. Numbers refer to $n$ described in the text.}
\end{figure}

The geometric feature with the greatest impact on $T(E)$ is $n$, the total number of Au nearest neighbors of the two binding sites. The right tail of the resonance peak, centered at $\sim$ -1 eV, is well fit to a Lorentzian form $T(E) = A/((E - E_{0})^{2} + (\Gamma/2)^{2})$, for each junction with $n$ $>$ 6 (Fig. 4(b)). For $n$ = 6, additional structure in T(E) appears near $E_{F}$. The relatively good fits for $n$ $>$ 6 reflect a modest electronic coupling between the BDA HOMO and an approximately constant density of states (DOS) in the leads. The structure in $T(E)$ near $E_{F}$ for $n$ = 6 is due to additional features in the projected DOS on the Au contacts (Fig. 4(b) inset), which arise from the interaction of bulk and surface states with the $s$-like orbitals of the undercoordinated adatom.  Overall, the transmission resonance energy $E_{0}$ (-1.27 to -0.94 eV) and peak width $\Gamma$ (0.34 to 0.56 eV) exhibit modest variation.  Further, the small increase in peak width $\Gamma$ with coordination $n$ leads to an increase in conductance, evident from the comparison of the Lorentzian conductance (predicted from the fit) with the calculated conductance in Fig. 4(c). However, the extra structure in $T(E)$ for $n$ = 6 increases the conductance beyond the Lorentzian model and reduces the variation of conductance with $n$.

From the experimental and theoretical results, we can conclude that the local amine-Au bonding motif does not vary significantly across different junction geometries. Large changes in the molecular tilt angle and bonding configuration can be accommodated with relatively small variations in the Au-N-C bond angles (which range from 112$^\circ$ to 130$^\circ$) and N-Au bond lengths (which range from 2.38 to 2.55 $\text{\AA}$). Because of the intrinsic isotropy of the Au $6s$ orbital, different amine-Au bond orientations have minimal impact on the electronic coupling. We note that junction-elongation process does not add significantly to the spread in conductance (Fig. 1). Most of the measured conductance steps are relatively short, consistent with breaking the N-Au bond \cite{Kosov07}. The well-defined amine-Au bonding motif, taken together with electronic flexibility of the amine-Au bond, result in a narrow conductance distribution across distinct junctions (Fig. 1).

Having developed a consistent picture for junction structure in agreement with experiment, we may provide a direct assessment of the DFT framework for computing conductance. Our calculations overestimate the conductance by $\sim$7$\times$. Given the relatively small effects of junction geometry on conductance, this discrepancy is unlikely to arise from finite temperature fluctuations.  Instead, the deviation likely originates from the use of a mean-field potential in DFT.  Indeed, many-electron interactions are essential to accurately position frontier molecular resonance energies in a junction, and the DFT errors can be large \cite{Neaton06}. Direct treatment of such interactions in transport through molecules is challenging and to date, studies have been confined to model systems \cite{Toher05,Delaney04,Ferretti05,Darancet07,Koentopp06,Sai05}. However, many-electron interactions are expected to reduce the DFT conductance, shifting the molecular resonances away from $E_{F}$ \cite{Darancet07,Koentopp06}, and narrowing the transmission peak widths in the weak coupling limit \cite{Ferretti05,Koentopp06}.  To have a sense for the magnitude of the changes required, either reduction of the width of the resonance $\Gamma$ from $\sim$0.5 to 0.2 eV or shift of the position of the resonance ($E_{0}$) from -1 eV to -3 eV relative to $E_{F}$, would bring the calculated conductance into line with the measured value. On physical grounds, the resonance position $E_{0}$ is probably too shallow because of self-energy errors inherent to the use of LDA or GGA Kohn-Sham eigenvalues as quasiparticle energies \cite{Hybertsen86}. A more complete first-principles treatment of electron correlation effects  \cite{Hybertsen86} suggests that the resonance should be deeper \cite{Neaton06}. Our simple estimate, including electrode polarization in a physically-motivated image charge model \cite{Neaton06}, suggests that a resonance position of -3 eV is plausible \cite{homo}.  Since the low bias conductance is determined by transmission through the tail of the HOMO-derived resonance, these shifts significantly alter the conductance, but they minimally affect the other calculated trends.

\begin{acknowledgments}
Portions of this work were performed at U.S. Department of Energy BES User Facilities the Molecular Foundry, LBNL (DE-AC0205CH11231)  and the Center for Functional Nanomaterials, BNL (DE-AC0298CH10886). This work was also supported by the Nanoscale Science and Engineering Initiative of the NSF (Award Numbers CHE-0117752 and CHE-0641532), the New York State Office of Science, Technology, and Academic Research and by NSF Grant No. DMR04-39768. 
\end{acknowledgments}


\begin{thebibliography}{99}
\bibitem{algo}
The algorithm locates peaks above a threshold of 5000 in the derivative of the log of the conductance trace. If the average conductance between two successive peaks is between 0.003 G$_{0}$ and 0.02G$_{0}$, the trace is considered to have a molecular step. Step length, slope and deviation are determined from a linear fit through the region.
\bibitem{Au}
Our bulk Au bandstructure matches plane wave calculations, and the bulk modulus (-7.5\%), lattice constant (+3.5\%), Au$_{2}$ vibrational frequency (-2.6\%) and Au(111) work function (+2.3\%) compare well to experiment.
\bibitem{homo}
The GGA gas-phase HOMO of BDA differs from the measured ionization potential (IP) by 3.1 eV. Electrostatic polarization effects from the Au electrodes are expected to reduce the IP \cite{Neaton06}. Using an image point-charge model, and assuming the image plane is $\sim$4 $\text{\AA}$ from the additional hole at the phenyl ring center, we estimate the IP will be reduced by $\sim$1 eV. We therefore estimate a net self-energy correction of $\sim$-2 eV for the BDA HOMO in the junction.
\end{thebibliography}
\end{document}